\documentclass[nofootinbib,preprint,superscriptaddress,showpacs,preprintnumbers,amsmath,amssymb,pra,aps]{revtex4-1}

\usepackage{amsmath,amsfonts,amssymb}
\usepackage{graphicx}

\usepackage{color}

\DeclareMathOperator{\arccosh}{arcosh}
\DeclareMathOperator{\arcsinh}{arsinh}

\begin{document}

\title{Semiclassical two-step model for strong-field ionization}

\author{N. I. Shvetsov-Shilovski}
\email{n79@narod.ru}
\affiliation{Institut f\"{u}r Theoretische Physik and Centre for Quantum Engineering and Space-Time Research, Leibniz Universit\"{a}t Hannover, D-30167 Hannover, Germany, EU}

\author{M. Lein}
\affiliation{Institut f\"{u}r Theoretische Physik and Centre for Quantum Engineering and Space-Time Research, Leibniz Universit\"{a}t Hannover, D-30167 Hannover, Germany, EU}

\author{L. B. Madsen}
\affiliation{Department of Physics and Astronomy, Aarhus University, 8000 {\AA}rhus C, Denmark, EU}

\author{E. R\"{a}s\"{a}nen}
\affiliation{Department of Physics, Tampere University of Technology, FI-33101 Tampere, Finland, EU}

\author{C. Lemell}
\affiliation{Institute for Theoretical Physics, Vienna University of Technology, A-1040 Vienna, Austria, EU}

\author{J. Burgd\"{o}rfer}
\affiliation{Institute for Theoretical Physics, Vienna University of Technology, A-1040 Vienna, Austria, EU}
\affiliation{Institute of Nuclear Research of the Hungarian Academy of Sciences, H-4001 Debrecen, Hungary, EU}

\author{D. G. Arb\'o}
\affiliation{Institute for Astronomy and Space Physics, IAFE (UBA-Conicet), Buenos Aires, Argentina}

\author{K. T\H{o}k\'esi}
\affiliation{Institute for Nuclear Research, Hungarian Academy of Sciences, H-4001 Debrecen, Hungary, EU}
\affiliation{ELI-HU Nonprofit Ltd., Dugonics t\'er 13, 6720 Szeged, Hungary, EU}

\date{\today}

\begin{abstract}
We present a semiclassical two-step model for strong-field ionization that accounts for path interferences of tunnel-ionized electrons in the ionic potential beyond perturbation theory. Within the framework of a classical trajectory Monte-Carlo representation of the phase-space dynamics, the model employs the semiclassical approximation to the phase of the full quantum propagator in the exit channel. By comparison with the exact numerical solution of the time-dependent Schr\"odinger equation for strong-field ionization of hydrogen, we show that for suitable choices of the momentum distribution after the first tunneling step, the model yields good quantitative agreement with the full quantum simulation. The two-dimensional photoelectron momentum distributions, the energy spectra, and the angular distributions are found to be in good agreement with the corresponding quantum results. Specifically, the model quantitatively reproduces the fan-like interference patterns in the low-energy part of the two-dimensional momentum distributions as well as the modulations in the photoelectron angular distributions.
\end{abstract}

\pacs{32.80Fb, 32.80Rm, 32.80 Wr}

\keywords{above-threshold ionization, semiclassical model, quantum interference} 

\maketitle

\section{Introduction}

Strong-field physics is concerned with highly nonlinear phenomena originating from the interaction of strong laser radiation with atoms and molecules. Above-threshold ionization (ATI), high-order harmonic generation (HHG), and non-sequential double ionization (NSDI) are the most well-known examples (see Refs.\ \cite{KrainovBook, BeckerRev02, MilosevicRev03, FaisalRev05, FariaRev11} for reviews). Among the main theoretical approaches used to understand these diverse phenomena are the direct numerical solution of the time-dependent Schr\"{o}dinger equation (TDSE), the strong-field approximation (SFA) \cite{Keldysh64, Faisal73, Reiss80}, and semiclassical models applying a classical description of an electron after it has been released from an atom, e.g., by tunneling ionization \cite{Dau, PPT, ADK}. The most widely known examples of semiclassical approaches are the two-step model for ionization \cite{Linden88, Gallagher88, Corkum89} and the three-step models for harmonic radiation and rescattering \cite{Kulander_Schafer93, Corkum93}. In the first step of the two-step model an electron tunnels out of an atom, and in the second step it propagates in the laser field. The third step involves the rescattering of the returning electron with the residual ion. Thus the three-step model allows for a qualitative description of rescattering-induced processes: high-order ATI, HHG, and NSDI.

Although significant progress has been made over the last two decades in development of the theoretical approaches based on the SFA and, particularly, on the TDSE (see, e.g., Refs.\ \cite{Muller99, Bauer06, Madsen07} and references therein), the semiclassical models are still extensively used in strong-field physics. The reason is that these models have a number of advantages. Indeed, semiclassical simulations can help to identify the specific mechanisms responsible for the phenomena under consideration, and provide an illustrative picture in terms of classical trajectories. For example, the three-step model explained the cutoffs in HHG \cite{Krause92, Lewenstein94} and high-order ATI spectra \cite{Paulus94}, the maximum angles in the photoelectron angular distributions \cite{Paulus94a}, and the characteristic momenta of recoil ions of the NSDI \cite{Faria03, Milosevic04}.

In their original formulation, the two-step and three-step models do not take into account the effect of the Coulomb potential of the parent ion on the electron motion after ionization. The inclusion of the Coulomb potential into the two-step model allowed to reveal the so-called Coulomb focusing effect \cite{BrabecIvanov96}. Employing classical trajectory Monte-Carlo (CTMC) simulations for the second step, the Coulomb cusp in the angular distribution of strong-field ionized electrons could be identified \cite{Dimitriou04}. Among the more recent examples of application of the semiclassical models with the Coulomb potential are the investigation of the so-called ``ionization surprise" \cite{Blaga09}, i.e., the low-energy structures in strong-field ionization by midinfrared pulses \cite{Quan09, Liu10, Yan10, Kastner12, Lemell12, Lemell13, Wolter14, Becker14, Dimitrovski15}, the study of the angular shifts of the photoelectron momentum distributions in close to circularly polarized laser fields \cite{Keller12, Maurer12, Shvetsov12, Maurer14}, and the analysis of the nonadiabatic effects in strong-field ionization (see, e.g., Refs.\ \cite{Boge13, Hofmann14, Geng14}). Semiclassical simulations are often computationally much simpler than the direct numerical solution of the TDSE. There are still many strong-field problems, for which semiclassical models are the only feasible approach. Well known examples of the latter category include the NSDI of atoms by elliptically \cite{Shvetsov08, XuWang09, Hao09} and circularly \cite{Uzer10} polarized fields as well as the NSDI of molecules \cite{Agapi10}. Therefore, improvements of the semiclassical models of strong-field phenomena are being sought with the goal to render them quantitatively predictive.

Recently, some progress among these lines has been achieved. For example, a criterion of applicability of the two-step model with the Coulomb potential of the parent ion was formulated in Ref.\ \cite{Shvetsov13}. Within a purely classical treatment of the electron dynamics subsequent to tunnel ionization, interference structures in the photoelectron spectrum and two-dimensional momentum distribution \cite{Paulus12,Arbo06,Moshammer09,Arbo10,Arbo14} cannot be reproduced. This deficiency has been overcome by a semiclassical model denoted by the authors of \cite{Li14} as the ``quantum trajectory Monte-Carlo (QTMC)'' \footnote{This model should not be confused with another approach termed quantum trajectory Monte-Carlo (QTMC) that was used for the solution of the Liouville equation for open quantum systems \cite{Minami03}. It is based on an ensemble of solutions of a stochastic Schr\"odinger equation each of which correspond to a quantum trajectory in Hilbert space.}. This model allows to include interference effects into the two-step model with the Coulomb potential. Accordingly, each classical trajectory is associated with a phase determined by the classical action, and the contributions of all trajectories leading to a given final momentum calculated by a CTMC approach are added coherently. The QTMC model has already been used in the study of nonadiabatic effects in tunneling ionization of atoms in elliptically polarized laser fields \cite{Geng14}. A similar approach, but disregarding the Coulomb potential, was used in Ref.~\cite{Li15} to investigate the holographic interference patterns in strong-field ionization of N$_2$, O$_2$, and CO$_2$. Very recently, the QTMC model has been applied to the identification of resonance structures in the low-energy part of the photoelectron spectra~\cite{Shao15} and to the study of the nonadiabatic subcycle electron dynamics in orthogonally polarized two-color laser fields \cite{Geng15}.

The Coulomb-corrected strong-field approximation (CCSFA) \cite{Yan10,Popruzhenko08,Popruzhenko08PRA} has been applied to analyze results of experiment and theory. The CCSFA invokes first-order perturbation theory \cite{PPT3} to include the Coulomb potential. Likewise, according to the Supplementary Material in Ref.~\cite{Li14}, the QTMC model includes the Coulomb effect in a perturbative manner; a point that we will discuss below. It is, therefore, of interest to formulate a semiclassical two-step (SCTS) model that accounts for the Coulomb potential beyond the semiclassical perturbation theory. Our present approach is based on the theory of semiclassical time-dependent propagators (see, e.g., Ref.\ \cite{TannorBook} for a text-book treatment). Here we derive a semiclassical expression for the transition amplitude for strong-field ionization that differs from the one used in the QTMC and CCSFA models improving the agreement with full quantum simulations.
 
The paper is organized as follows. In Sec.~\ref{sec2} we briefly review previous two-step models that invoke semiclassical approximations at various stages of their development. In Sec.~\ref{sec3} we present our semiclassical two-step model that combines the CTMC method for trajectory sampling with the phase of the semiclassical propagator and discuss its numerical implementation. The application to the benchmark case of strong-field ionization of atomic hydrogen, and the comparison with TDSE results are presented in Sec.~\ref{sec4} followed by concluding remarks in Sec.~\ref{sec5}. Atomic units are used throughout the paper unless indicated otherwise. 

\section{Two-step models}\label{sec2}
The two-step models for direct strong-field ionization as well as their three-step extensions typically invoke semiclassical approximations to the full quantum dynamics at various levels. We briefly sketch the major steps involved in order to delineate the point of departure of the present SCTS model. It should be stressed that our model is different from SFA-type models such as the CCSFA model \cite{Yan10,Popruzhenko08,Popruzhenko08PRA} as the latter are applicable for arbitrary values of the Keldysh parameter $\gamma = \omega\kappa/F$ \cite{Keldysh64} (here $\omega$ is the angular frequency of the laser field, $F$ is the field amplitude, and $\kappa=\sqrt{2I_p}$, where $I_p$ is the ionization potential). In contrast to this, we employ instantaneous tunneling ionization rates. The description of the ionization step by tunneling is expected to be accurate only for small values of the Keldysh parameter, $\gamma\ll 1$.

The starting point of the semiclassical approximation is the assumption that the (classical) action in the Feynman propagator is asymptotically large compared to the quantum action $\hbar$ such that the path integral over (in general, non-classical) paths can be performed by the saddle-point approximation. Equivalently, the semiclassical approximation (SCA) can be viewed as the leading term in an $\hbar^n$ expansion as $\hbar\to 0$. Accordingly, the expression for the matrix element of the semiclassical propagator $U_{SC}$ between the initial state at time $t_1$ and the final state at time $t_2 \left(t_2>t_1\right)$ reads  \cite{Miller71,Walser03,Spanner03} (see Refs.~\cite{GrossmannBook} and \cite{TannorBook} for a text-book treatment):
\begin{equation}
\left\langle q_2\right|U_{\rm{SC}}\left(t_2, t_1\right)\left|q_1\right\rangle=\left[-\frac{\partial^2\phi_{1}\left(q_1,q_2\right)/\partial q_1 \partial q_2}{2\pi i}\right]^{1/2}\exp\left[i \phi_{1}\left(q_1,q_2\right)\right]. 
\label{Matrixelem1}
\end{equation}
Here $q_1$ and $q_2$ are the spatial coordinates of a particle at times $t_1$ and $t_2$, respectively, and the phase $\phi_{1}\left(q_1,q_2\right)=S_1(q_1,q_2)/\hbar$ is given in terms of the action $S_1$,
\begin{equation}
S_{1}\left(q_1,q_2\right)=\int_{t_1}^{t_2}\left\{p\left(t\right)\dot{q}\left(t\right)-H\left[q\left(t\right),p\left(t\right)\right]\right\}dt,
\label{Phase1}
\end{equation}
where, in turn, $H\left[q\left(t\right),p\left(t\right)\right]$ is the classical Hamiltonian function as a function of the canonical coordinates $q(t)$ and momenta $p(t)$. For simplicity, we use in Eqs.~(\ref{Matrixelem1},\ref{Phase1}) a notation for 1D systems. Our applications in the following account, however, for the full dimensionality of the problems. The prefactor in Eq.~(\ref{Matrixelem1}), frequently referred to as van the Vleck (vV) determinant for multi-dimensional systems is independent of $\hbar$. The phase factor $\exp (i\phi_i)$ is non-analytic as $\hbar\to 0$ and accounts for the non-uniform approach to the classical limit via increasingly fast oscillations (see Ref.~\cite{Yoshida04}).

For classically allowed processes, the square modulus of the vV-determinant gives the classical phase-space density (or probability density) for the phase flow from $q_1$ to $q_2$ within the time interval $t_2-t_1$. Higher order corrections in $\hbar$ are neglected in Eq.~(\ref{Matrixelem1}) from the outset. Using atomic units in the following we will not display the $\hbar$ dependence explicitly but, instead, express the semiclassical limit in terms of the de Broglie wavelength $\lambda_{dB}$ exploiting the equivalence of vanishing de Broglie wavelength $\lambda_{dB}\to 0$ and the limit $\hbar\to 0$.

The point to be emphasized is that the applicability of the saddle-point approximation, and, in turn, the semiclassical limit for the ionization process in the presence of laser and Coulomb fields is, \textit{a priori}, not obvious. The tunneling process is intrinsically non-classical and the de Broglie wavelength $\lambda_{dB}(E)$ of slow electrons close to the tunneling exit is, in general, not small compared to the exit coordinate $\eta$ or the width of the barrier, i.e., the semiclassical relation $\lambda_{dB}\ll\eta$ is violated.

\subsection{SCA for the first step}
The starting point for the first step of strong-field ionization, the tunneling through the barrier formed by the atomic (ionic) potential and the interaction with the electromagnetic field, is the quantum transition matrix element in distorted-wave Born approximation (SFA)
\begin{equation}
M_\mathrm{SFA}\left(\vec p\right)=-i\int_{-\infty}^\infty dt\, \langle\psi_{\vec p}(t)|V_{L}(t)|\psi_i\left(t\right)\rangle\, ,\label{MSFA}
\end{equation}
where $|\psi_i\left(t\right)\rangle$ is the bound initial state and $|\psi_{\vec{p}}(t)\rangle$ is the Volkov state after tunneling,
\begin{equation}
\psi_{\vec{p}}(\vec r, t)=\exp\left[i\left(\vec p + \vec{A}\left(t\right)/c\right)\vec r-\frac{i}{2}\int_{-\infty}^t dt'\,\left(\vec p +\vec A(t')/c\right)^2\right]\, .
\end{equation}
Eq.~(\ref{MSFA}) is referred to as the strong-field approximation as in the final state the ionic potential is considered to be negligible in comparison to the interaction $V_{L}\left(t\right)=\vec{F}\left(t\right)\cdot \vec{r}$ with the strong electric field $\vec{F}\left(t\right)=-\frac{1}{c}\frac{d\vec{A}}{dt}$. The time integral is evaluated within the framework of the saddle-point approximation assuming that the effective phase (or action)
\begin{equation}
S(\vec p,t)=\int_{-\infty}^{t} dt'\, \left\{\frac{1}{2}[\vec p+\vec A(t')/c]^2-\varepsilon_i\right\}
\end{equation}
is large and rapidly varying with $t$, thereby invoking the semiclassical (SC) limit. The prerequisites for the applicability of this semiclassical limit are the ponderomotive energy $U_p$ and the ionization potential $I_p=-\varepsilon_i$ to be large compared to the photon energy $\omega$, $U_p/\omega \gg 1$ and $I_p/\omega\gg 1$. Unlike Eq.~(\ref{Matrixelem1}), the semiclassical approximation is applied to the transition matrix element [Eq.~(\ref{MSFA})] rather than to the propagator. Accordingly (see, e.g., \cite{Milosevic06,Yan12}),
\begin{equation}
M_\mathrm{SFA}^\mathrm{SC}=\sum_j\frac{\exp [iS(\vec p,t_s^j)]}{\left[\frac{\partial^2}{\partial t^2} S(\vec p,t_s^j)\right]^{1/2}}\, V_{\vec p}^j\label{MSFASC}
\end{equation}
with $V_{\vec p}^j$ containing the spatial dependencies of the transition matrix element. The saddle point equation
\begin{equation}
\frac{\partial}{\partial t} S(\vec p,t_s)=\frac{1}{2}(\vec p+\vec A(t_s)/c)^2+I_p=0\label{SPE}
\end{equation}
has complex solutions in $t$, $t_s^j=t_0^j+it_t^j$ where the real part $t_0$ is referred to as the ionization time and the imaginary part $t_t$ as the tunneling time. Because of the complex solutions, the emerging trajectories are often referred to as quantum trajectories, see Refs.~\cite{Salieres01} and \cite{Milosevic06}. When more than one saddle point contributes ($j=1,...$), Eq.~(\ref{MSFASC}) can give rise to semiclassical path interferences.

Frequently employed approximate evaluations of Eq.\ (\ref{MSFASC}) include the Perelomov-Popov-Terent'ev (PPT) or Ammosov-Delone-Krainov (ADK) rate for adiabatic tunneling in which $t_0$ coincides with the extremum of the electric field $F$ \cite{Dau, PPT, ADK}:
\begin{equation}  
\label{tunrate}
w\left(t_{0},v_{0, \perp}\right)\sim\left|M_\mathrm{SFA}^\mathrm{SC}\right|^2\sim\exp\left(-\frac{2\kappa^3}{3F\left(t_0\right)}\right)\exp\left(-\frac{\kappa v_{0,\perp}^{2}}{F\left(t_0\right)}\right).
\end{equation} 
For simplicity, we omit the preexponential factor in Eq.~(\ref{tunrate}). Although this factor changes the total ionization rate by several orders of magnitude, for atoms it only slightly affects the shape of the photoelecton momentum distributions. When applying Eq.~(\ref{tunrate}), a simple and frequently used choice is that the electron emerges with vanishing velocity component along the laser polarization direction $v_{0,z}=0$ while $v_{0,\perp}$ is Gaussian distributed. It is common to apply Eq.~(\ref{tunrate}) as a quasistatic rate \cite{Yudin01}, i.e., for tunneling ionization for laser phases other than the field extremum, with $F\left(t_0\right)$ in Eq.~(\ref{tunrate}) denoting the instantaneous field. Non-zero longitudinal velocity components $v_{0,z}\ne 0$ appear near the tunnel exit when the sub-barrier motion is modeled by the strong-field approximation \cite{Li16}.

The coordinates of the tunneling exit can be conveniently determined by using the fact that for an electron in a time-independent electric field $F$ and the Coulomb potential, $-Z/r$, both the classical Hamilton-Jacobi equation \cite{Born25} and the stationary Schr\"odinger equation are separable in parabolic coordinates (see Refs.~\cite{Schrodinger} and \cite{Dau})
\begin{eqnarray}
\label{parab}
\xi=r+z,~ & \eta=r-z,~ &\phi=\arctan\left(\frac{y}{x}\right)\, .
\end{eqnarray} 
If the electric field points along the positive $z$ direction, the electron is trapped by an attractive potential along the $\xi$ coordinate and can tunnel out only in the $\eta$ direction. The tunnel exit coordinate $\eta$ is then obtained from the equation
\begin{equation}
\label{tunexit}
-\frac{\beta_{2}\left(F\right)}{2\eta}+\frac{m^2-1}{8\eta^2}-\frac{F\eta}{8}=-\frac{I_{p}\left(F\right)}{4},
\end{equation}  
where 
\begin{equation}
\label{const}
\beta_{2}\left(F\right)=Z-\left(1+\left|m\right|\right)\frac{\sqrt{2I_{p}\left(F\right)}}{2}
\end{equation}
is the separation constant (see, e.g., Ref.~\cite{Bisgaard04}) and $m$ is the magnetic quantum number of the initial state. In Eq.~(\ref{const}) we have allowed for the Stark shift of the initial state, i.e., of the ionization potential:
\begin{equation}
\label{ipstark}
I_{p}\left(F\right)=I_{p}+\frac{1}{2}\left(\alpha_{N}-\alpha_{I}\right)F^{2},
\end{equation} 
where $F$ is the instantaneous field amplitude at the ionization time $t_0$ and $\alpha_{N}$ and $\alpha_{I}$ are the static polarizabilities of an atom and of its ion, respectively \cite{Dimitrovski10}. With Eqs.~(\ref{tunexit}) and (\ref{tunrate}), the initial conditions for the propagation of trajectories subsequent to tunneling ionization, i.e., for the second step, are determined.

As the focus of the present work is the improved semiclassical description of the second step, we treat in the following the output of the first step, in particular the initial velocity (or momentum) distribution at the tunneling exit [see, e.g., Eq.~(\ref{tunrate})] as adjustable input. We will use two different initial phase-space distributions resulting from the tunneling step as initial conditions for the post-tunneling semiclassical propagation. Both choices of distributions are described by Eq.~(\ref{tunrate}). The difference is that in one case, the initial parallel velocity $v_{0,z}$ is set to zero, whereas in the other case, it is set to a nonzero value predicted by the SFA \cite{Li16}.   

\subsection{SCA for the second step}
The position and momentum distributions at the tunneling exit serve as initial conditions for the propagation of classical trajectories in the second step. In the simplest approximation, the quiver motion for a selected set of trajectories for free electrons in the electromagnetic field is treated thereby neglecting the atomic force field \cite{Corkum89, Kulander_Schafer93, Corkum93, Lewenstein94}. More advanced descriptions employ full CTMC simulations treating the laser field and the atomic force field on equal footing by solving Newton's equation of motion
\begin{equation}
\label{newton}
\frac{d^2\vec{r}(t)}{dt^2} = - \vec{F}(t) - \frac{Z\vec{r}(t)}{r^3(t)}
\end{equation} 
for a large number of initial conditions (typically $\ge 10^7$) thereby sampling the initial phase-space distribution after tunneling \cite{Dimitriou04}.

For the propagation of the electron in the combined fields [Eq.~(\ref{newton})], a semiclassical approximation in terms of a coherent superposition of amplitudes appears justified, since classical-quantum correspondence holds separately for both the propagation in the Coulomb field and in the laser field. For the linear potential of a charged particle in the external field, $V_{L}=\vec{F}\cdot\vec{r}$, the Ehrenfest theorem holds for Newton's law, $\langle\vec\nabla V_{L}\rangle=\vec\nabla V_{L}(\langle\vec r\rangle)$. For the long-range Coulomb potential $\lambda_{dB}(E)$ is negligible compared to the infinite range of the potential at any energy $E$. However, an extension of CTMC simulations to the semiclassical domain faces considerable difficulties in view of the intrinsic numerical instability which is closely related to the non-uniform convergence to the classical limit mentioned above. Superposition of a large number of amplitudes associated with trajectories with rapidly oscillating phases fails to yield converged scattering amplitudes in the asymptotic limit $t\to\infty$ \cite{Yoshida04}.

One key ingredient is therefore the binning of the trajectories according to the appropriate final canonical momenta and restricting coherent superpositions to those trajectories within each bin. For bound state excitation driven by ultrashort pulses this corresponds to binning of the action variable, i.e., to a quantization of classical trajectories. This quantized classical trajectory Monte-Carlo method \cite{Reinhold09} can accurately account for quantum revivals and dephasing in Rydberg manifolds.

For strong-field ionization, the final states lie in the continuum and are binned according to their momenta in cells in momentum space \cite{Geng14,Geng15}, $[p_i,p_i+\Delta p_i]$, with $i=x,y,z$. Accordingly, the amplitudes associated with all $n_p$ trajectories taking off at $t_0^j$ with initial velocity $\vec{v}_0^j$ ($j=1,...,n_p$) reaching the same bin centered at $\vec{p}=\left(p_x,p_y,p_z\right)$ are added coherently. Thus the ionization probability $R\left(\vec{p}\right)$ for this final momentum $\vec{p}$ is given by
\begin{equation}
\label{prob}
R\left(\vec{p}\right)=\left|\sum_{j=1}^{n_p}\sqrt{w\left(t_{0}^{j},\vec v_0^{j}\right)}\exp\left[i\Phi\left(t_{0}^{j},\vec v_0^{j}\right)\right]\right|^{2},
\end{equation}  
where $w(t_{0}^{j},v_0^{j})$ is the probability density of the initial conditions. The sum over $j$ samples the classical phase flow from $\vec v_0$ to the bin $\vec p$ corresponding to the vV determinant as determined by CTMC. $\Phi (t_{0}^{j},\vec v_0^{j})$ is the phase that each trajectory carries. When the interference phases of trajectories reaching the same bin are neglected, the classical CTMC probability density
\begin{equation}
\label{prob0}
R\left(\vec{p}\right)=\sum_{j=1}^{n_p}w\left(t_{0}^{j},\vec v_{0}^{j}\right)
\end{equation}
emerges. In the QTMC model \cite{Li14}, the phase in Eq.~(\ref{prob}) was approximated by
\begin{equation}
\Phi^{\rm{QTMC}}\left(t_{0}^{j},\vec v_{0}^{j}\right)\approx I_{p}{t_{0}^{j}}-\int_{t_{0}^{j}}^\infty \left(\frac{v^2(t)}{2}-\frac{Z}{r(t)}\right)\, dt\, .\label{phaseQTMC}
\end{equation}
We will relate the phase in Eq.~(\ref{phaseQTMC}) to our semiclassical phase in Sec.~III~A.

\section{Formulation of the model}\label{sec3}

\subsection{Semiclassical expression for the phase}
The two key ingredients of the present semiclassical two-step model are the choice of an initial momentum distribution emerging from the first tunneling step based on SFA estimates and a proper semiclassical description for the second step. This approach accounts for the expectation that the semiclassical limit is applicable for the evolution of the liberated electron in the combined laser and ionic force fields. We describe the second step of the two-step model using the expression for the matrix element of the semiclassical propagator $U_{\rm{SC}}$. In addition to its coordinate representation [Eq.~(\ref{Matrixelem1})] three equivalent forms involving different combinations of phase-space coordinates exist \cite{Miller71}
\begin{subequations}
\begin{align}
\label{Matrixelem2}
\left\langle q_2\right|U_{\rm{SC}}\left(t_2,t_1\right)\left|p_1\right\rangle & = \left[-\frac{\partial^2\phi_{2}\left(p_1,q_2\right)/\partial p_1 \partial q_2}{2\pi i}\right]^{1/2}\exp\left[i \phi_{2}\left(p_1,q_2\right)\right]\, ,\\ 
\label{Matrixelem3}
\left\langle p_2\right|U_{\rm{SC}}\left(t_2,t_1\right)\left|q_1\right\rangle & = \left[-\frac{\partial^2\phi_{3}\left(q_1,p_2\right)/\partial q_1 \partial p_2}{2\pi i}\right]^{1/2}\exp\left[i \phi_{3}\left(q_1,p_2\right)\right]\, ,\\
\label{Matrixelem4}
\left\langle p_2\right|U_{\rm{SC}}\left(t_2,t_1\right)\left|p_1\right\rangle & = \left[-\frac{\partial^2\phi_{4}\left(p_1,p_2\right)/\partial p_1 \partial p_2}{2\pi i}\right]^{1/2}\exp\left[i \phi_{4}\left(p_1,p_2\right)\right]\, .
\end{align}
\end{subequations}
They describe the propagation from the initial position $(q_1)$ or momentum coordinate $(p_1)$ to a final position $(q_2)$ or momentum coordinate $(p_2)$ within the time interval $t_2-t_1$. The phases $\phi_i,\, i=2,3,4$ in Eqs.~(\ref{Matrixelem2}-c) are given by the classical action associated with the corresponding canonical transformations
\begin{subequations}
\begin{align}
\label{Phase2}
\phi_2\left(p_1,q_2\right)&=\phi_{1}\left(q_1,q_2\right)+p_{1}q_{1}\, ,\\
\label{Phase3}
\phi_3\left(q_1,p_2\right)&=\phi_{1}\left(q_1,q_2\right)-p_{2}q_{2}\, ,\\
\label{Phase4}
\phi_4\left(p_1,p_2\right)&=\phi_{1}\left(q_1,q_2\right)+p_{1}q_{1}-p_{2}q_{2}\, ,
\end{align}
\end{subequations}
with $\phi_1(q_1,q_2)$ given by Eq.~(\ref{Phase1}). The generalization of Eqs.~(\ref{Matrixelem2}-c) and (\ref{Phase2}-c) to three dimensions is straightforward: $q_1$, $q_2$, $p_1$, and $p_2$ should be replaced by the corresponding vectors $\vec{r}_1$, $\vec{r}_2$,  $\vec{p}_1$, and $\vec{p}_2$ \cite{Miller71, Kay2005}. The products $p_{1}q_{1}$ and $p_{2}q_{2}$ in Eqs.~(\ref{Phase2}-c) are to be replaced by the corresponding scalar products: $\vec{p}_1\cdot\vec{r}_1$ and $\vec{p}_2\cdot\vec{r}_2$.

It is now of interest to inquire which of the propagator matrix elements is appropriate for the second step of strong-field ionization. Semiclassical scattering characterized by a transition from momentum $\vec p_1$ at $t\to-\infty$ to $\vec p_2$ at $t\to\infty$ is described by the propagator Eq.~(\ref{Matrixelem4}) with the compensated action $\phi_4$ given by [Eq.~(\ref{Phase4})]:
\begin{equation}
\label{phaseACT}
\phi_{4}\left(\vec{p}_1,\vec{p}_2\right)=\int_{t_1}^{t_2}\left\{-\vec{r}\left(t\right)\cdot\dot{\vec{p}}\left(t\right)-H\left[\vec{r}\left(t\right),\vec{p}\left(t\right)\right]\right\}dt\, .
\end{equation}
For strong field ionization representing a half-scattering process of an electron initially located near the nucleus and emitted with final momentum $\vec p_2(t\to\infty)$, the propagator Eq.~(\ref{Matrixelem3}) with action $\phi_3$ should be applicable for trajectories launched with initial phase $\exp\left(iI_pt_{0}\right)$ according to the time evolution of the ground state. This choice is based on the assumption of well localized starting points in coordinate space $\vec{r}_{1}$ near the tunnel exit [Eq.~(\ref{tunexit})] with negligible phase accumulation under the barrier in position-space representation.  We have 
\begin{equation}
\phi_3(\vec r_1,\vec p_2) = \phi_4(\vec p_1,\vec p_2)-\vec p_1\cdot\vec r_1
\end{equation}
Note that in the limit of vanishing longitudinal velocity at the tunneling exit ($v_{0,z}=0$), $\vec p_1$ is orthogonal to $\vec r_1$ and, hence, $\phi_3$ and $\phi_4$ coincide. In the following we include the phase contribution $\vec p_1\cdot\vec r_1$ for non-zero $v_{0,z}$. Its contribution to the interference pattern discussed below is found, however, to be of numerically minor importance. 

In our model we restrict ourselves to exponential accuracy. Thus we ignore the preexponential factor of the matrix element. Using $\phi_3(\vec r_1,\vec p_2)$ in Eq.~(\ref{prob}) yields the semiclassical approximation for the probability for strong-field ionization with final momentum $\vec p$,
\begin{equation}
\label{prob_sim}
R(\vec p)=\left|\sum_{j=1}^{n_p}\sqrt{w\left(t_{0}^{j},\vec v_0^{j}\right)}\exp\left[i\Phi\left(t_{0}^{j},\vec v_0^{j}\right)\right]\right|^{2},
\end{equation}
with
\begin{eqnarray}
\Phi\left(t_{0}^{j},\vec v_0^{j}\right)&=& - \vec v_0^{j}\cdot\vec r_t(t_0^{j}) + I_{p}t_{0}^{j} - \int_{t_0^{j}}^\infty dt\, \left\lbrace\dot{\vec p}(t)\cdot\vec r(t)+H[\vec r(t),\vec p(t)]\right\rbrace \nonumber\\
&=& - \vec v_0^{j}\cdot\vec r_t(t_{0}^{j}) + I_{p}t_{0}^{j} - \int_{t_{0}^{j}}^\infty dt\, \left\lbrace\frac{v^2(t)}{2}-\frac{2Z}{r(t)}\right\rbrace\, .\label{phase}
\end{eqnarray}
The generalization of Eq.~(\ref{phase}) to an effective potential $V(\vec{r})$ reads
\begin{equation}
\Phi\left(t_{0}^{j},\vec v_0^{j}\right)= - \vec v_0^{j}\cdot\vec r_t(t_0^{j}) + I_{p}t_{0}^{j} - \int_{t_0^{j}}^\infty dt\, \left\lbrace\frac{v^2(t)}{2}+V[\vec{r}(t)]-\vec r(t)\cdot\vec\nabla V[\vec{r}(t)]\right\rbrace\, .
\label{Phi_sim}
\end{equation}
Equation~(\ref{Phi_sim}) is applicable to effective one-electron descriptions of ionization of multi-electron systems employing model or pseudopotentials \cite{Tong15}. It should be noted, however, that in the presence of a strong short-ranged contribution to $V(r)$ the validity of the underlying semiclassical approximation, $\lambda_{dB}\ll R$, where $R$ is the range of the short-ranged contribution, is not obvious and remains to be verified.

For the Coulomb potential $V\left(\vec{r}\right)=-Z/r$, the phase of the QTMC model can be obtained from Eq.~(\ref{Phi_sim}) by neglecting the term $\vec{r}\left(t\right)\cdot \vec{\nabla} V\left[\vec{r}\left(t\right)\right]$ in the integrand of Eq.~(\ref{Phi_sim}). Thus the SCA phase given by Eq.~(\ref{phase}) differs from that of the QTMC model [Eq.~(\ref{phaseQTMC})]: The Coulomb interaction enters with doubled weight. The factor 2 originates from properly accounting for elastic scattering in Eqs.~(\ref{Phase3}) and (\ref{Phi_sim}), i.e., from fully accounting for the Coulomb potential in the compensated action $\phi_4(\vec p_1,\vec p_2)$. Note, that this compensated action accounts for elastic scattering also in the absence of time-dependent processes. By contrast, Eq.~(\ref{phaseQTMC}) yields for any time-independent Hamiltonian only a trivial trajectory-independent phase $\sim\int dt(H+I_p)$. The QTMC phase can therefore be viewed as an approximation to the full semiclassical phase Eq.~(\ref{phase}). 

\subsection{Numerical implementation}
In the presence of long-range interactions the calculation of the semiclassical transition amplitude [Eq.~(\ref{prob_sim})] for strong-field ionization requires special care in view of divergent phases and the large number of trajectories for a dense sampling of phase space needed for achieving sufficient resolution for the multi-differential ionization probability.

We subdivide the integration interval $[t_0^j,\infty]$ into two intervals $[t_0^j,\tau_f]$ and $[\tau_f,\infty]$ where $\tau_f$ is the time at which the laser pulse has concluded and beyond which the energy $H(\tau_f)=E$ is conserved along the outgoing Kepler hyperbola. For pure Coulomb potentials the asymptotic phase-space coordinates $[\vec p(\infty)]$ can be determined by the analytic Coulomb mapping of the coordinates $[\vec r(\tau_f),\vec p(\tau_f)]$ for given energy $E$,
\begin{equation}
\frac{p^2(\infty)}{2}=\frac{p^2(\tau_f)}{2}-\frac{1}{r(\tau_f)}\, ,
\end{equation}
the angular momentum
\begin{equation}
\vec L=\vec r(\tau_f)\times\vec p(\tau_f)
\end{equation}
and the Runge-Lenz vector
\begin{equation}
\vec a=\vec p(\tau_f)\times\vec L - \vec r(\tau_f)/r(\tau_f)\, .
\end{equation}
The asymptotic momentum follows from (see Ref.~\cite{Shvetsov12} that corrects the misprint in \cite{Shvetsov09}):
\begin{equation}
\label{mominf}
\vec{p}(\infty)=p(\infty)\frac{p(\infty)\left(\vec{L}\times\vec{a}\right)-\vec{a}}{1+p^2(\infty)L^2}\, .
\end{equation}
The phase Eq.~(\ref{phase}) can be analogously decomposed as
\begin{equation}
\Phi\left(t_{0}^{j},\vec v_0^{j}\right)= - \vec v_0^{j}\cdot\vec r_t(t_0^{j}) + I_{p}t_{0}^{j} - \int_{t_{0}^{j}}^{\tau_f} dt\, \left\lbrace\frac{v^2(t)}{2}-\frac{2Z}{r(t)}\right\rbrace-\int_{\tau_f}^{\infty} dt\, \left\lbrace E-\frac{Z}{r(t)}\right\rbrace\, .\label{phase_decom}
\end{equation}
We furthermore separate the last term in Eq.~(\ref{phase_decom}) representing the scattering phase accumulated in the asymptotic interval $[\tau_f,\infty]$ into the parts with time-independent and time-dependent integrand. The first part yields the linearly divergent contribution
\begin{equation}
\lim_{t\to\infty}E(t-\tau_f)\, .
\end{equation}
Since only the relative phase between those trajectories arriving in the same bin contribute to the probability (\ref{prob_sim}) whose final momenta and, therefore, energies coincide, the term $(E_j-E_{j'})(t-\tau_f)$ vanishes. This allows for the reduction of the integral for the interval $[\tau_f,\infty]$ to the Coulomb phase
\begin{equation}
\Phi_f^C(\tau_f)=\int_{\tau_f}^{\infty}\frac{dt}{r(t)}\label{phase_Coul}
\end{equation}
which is still divergent. The regularization of this integral can be performed by analytic Coulomb mapping for Kepler hyperbolae: the distance from the Coulomb center (i.e., from the ion) at a given time $t$ reads (see, e.g., Ref.~\cite{Dau1})
\begin{equation}
\label{distCoulomb}
r\left(t \right)=b\left(g\cosh\xi-1\right),
\end{equation}
where $b=1/\left(2E\right)$, $g=\sqrt{1+2EL^2}$, and the parameter $\xi=\xi\left(t\right)$ is determined from
\begin{equation}
\label{tparam}
t=\sqrt{b^3}\left(g\sinh{\xi}-\xi\right)+C.
\end{equation}
The constant $C$ in Eq.~(\ref{tparam}) can be found from the initial conditions for the motion in the Coulomb field, i.e., from the position $\vec{r}(\tau_f)$ and momentum $\vec p(\tau_f)$ of an electron at $t=\tau_f$. With the equations (\ref{distCoulomb}) and (\ref{tparam}) the integral in Eq.~(\ref{phase_Coul}) gives
\begin{equation}
\label{phaseinfres}
\Phi_{f}^C\left(\tau_f\right)=\sqrt{b}\left[\xi\left(\infty \right)-\xi\left(\tau_f\right)\right]. 
\end{equation}
Thus for every trajectory we need to calculate $\xi\left(\infty \right)$ and $\xi\left(\tau_f\right)$. Since $\xi\to\infty$ for $t\to\infty$, we can discard the decaying exponent in $\sinh\xi=\left[\exp\left(\xi\right)+\exp\left(-\xi\right)\right]/2$ and neglect both $C$ and $\xi$ compared to $\exp\left(\xi\right)$ in the asymptotic limit [Eq.~(\ref{tparam})]. Consequently, we find for asymptotically large $\xi$
\begin{equation}
t \approx\sqrt{b^3} g \exp(\xi)/2
\end{equation}
from which follows
\begin{equation}
\xi\left(t\to\infty\right)\approx\ln\left(\frac{2t}{g\sqrt{b^3}}\right).
\label{tparambig}
\end{equation}
Since we are interested in the relative phases of the interfering trajectories within the same bin and $b$ depends only on the energy $E$, we can disregard $2t/\sqrt{b^3}$ under the logarithm in Eq.~(\ref{tparambig}). Note that $g$ depends on both electron energy and angular momentum $L$. The latter is different for different interfering trajectories within a given bin of the momentum space. Thus we set $\xi\left(\infty\right)=-\ln\left(g\right)$. For the lower boundary in Eq.~(\ref{phaseinfres}) we find from Eq.~(\ref{distCoulomb})
\begin{equation}
\label{acosh}
\xi\left(\tau_f\right)=\pm\arccosh\left\{\frac{1}{g}\left[\frac{r\left(\tau_f\right)}{b}+1\right]\right\}, 
\end{equation}
where the sign still needs to be determined. Taking into account that $dr/dt=\vec{r}\vec{v}/r$ and $dr/dt=\left(dr/d\xi\right)/\left(dt/d\xi\right)$ and using Eqs.~(\ref{distCoulomb}) and (\ref{tparam}), we find for $\xi\left(\tau_{f}\right)$
\begin{equation}
\label{acosh2}
\xi\left(\tau_f\right)=\arcsinh\left\{\frac{\vec{r}(\tau_f)\cdot\vec{p}(\tau_f)}{g\sqrt{b}}\right\}.
\end{equation}
Thus the finite interference contribution from the Coulomb phase becomes
\begin{equation}
\Phi_f^C(\tau_f)=-\sqrt{b}\left[ \ln g+\arcsinh\left\{\frac{\vec{r}(\tau_f)\cdot\vec{p}(\tau_f)}{g\sqrt{b}}\right\}\right]\, .\label{phase_asym}
\end{equation}
We note that such asymptotic Coulomb phase contributions are missing in the QTMC model \cite{Li14}.

In order to achieve convergent semiclassical amplitudes based on Monte-Carlo sampling of a large number of classical trajectories, efficient sampling of initial conditions is essential. One standard method employs initial sets of $t_0$ and $\vec v_0$ that are either uniformly randomly distributed or distributed on a grid (e.g., in Ref.~\cite{Li14}). This results in sampling of a large number of trajectories with relatively small weights [see, e.g., Eq.~(\ref{tunrate})], which contribute to the final momentum distribution only to a small extent. Here, we implement an alternative Monte-Carlo algorithm based on importance sampling. We account for the importance of a given trajectory already at the sampling stage, i.e., before the integration of the equations of motion (\ref{newton}). Importance sampling is particularly significant in the presence of interference because typically many more trajectories are needed to resolve fine interference structures compared to CTMC simulations without interference (see Sec.~\ref{sec4}). Calculation of the ionization probability with trajectories selected by importance sampling is given [instead of Eq.~(\ref{prob_sim})] by
\begin{equation}
\label{prob1}
R\left(\vec{p}\right)=\left|\sum_{j=1}^{n_p}\exp\left[i\Phi\left(t_{0}^{j},\vec v_0^{j}\right)\right]\right|^2
\end{equation}
with ionization times $t_{0}^{j}$ and initial velocities $\vec v_0^{j}$ distributed according to the \textit{square root} of the tunneling probability, $\sqrt{w(t_0,\vec v_0)}$ [Eq.~(\ref{tunrate})]. Depending on the laser parameters and tunneling probabilities this importance sampling algorithm can significantly increase the computational speed and convergence as a function of the number of simulated trajectories.

\section{Results and discussion}\label{sec4}  

In our simulations we use a few-cycle linearly polarized laser pulse defined in terms of a vector potential that is present between $t=0$ and $t=\tau_{f}$,
\begin{equation} 
\label{vecpot}
\vec{A}\left(t\right)=(-1)^{n+1}\frac{cF_0}{\omega}\sin^{2}\left(\frac{\omega t}{2n}\right)\sin\left(\omega t\right)\vec{e}_z.
\end{equation}  
Here $\vec{e}_z$ is the unit vector pointing in polarization direction and $n$ is the number of optical cycles of the field with $\tau_{f}=2\pi n/\omega$. The electric field is obtained from Eq.~(\ref{vecpot}) by $\vec{F}\left(t\right)=-\frac{1}{c}\frac{d\vec{A}}{dt}$. We solve Newton's equations of motion using a fourth-order Runge-Kutta method with adaptive step size \cite{Numerical} and calculate the phase [Eq.(\ref{phase})] by adding an extra equation to the system of equations of motion.
 
Here we consider linearly polarized fields only. Because of the rotational symmetry with respect to the polarization direction of the laser pulse the semiclassical simulations for a linearly polarized field can be performed employing only two degrees of freedom $(z,r_\perp)$. This reduces the numerical complexity of the problem significantly. Indeed, in order to achieve convergence of the interference oscillations, we need about $10^{9}$ trajectories (for a comparison, $1.5\times10^{6}$ trajectories were sufficient in the CTMC simulation to calculate electron momentum distributions without interference \cite{Shvetsov14}). Nearly the same number of trajectories is used in CCSFA calculations (see, e.g., Ref.~\cite{Yan10}). Thus about 100 times more trajectories are needed for the semiclassical simulations when interference is included. Simulations of interactions with elliptically or circularly polarized laser fields will require an even larger number of trajectories. 

At the intensity of $9\times10^{13}$ W/cm$^2$ the size of the bin on the $\left(p_z,p_{\perp}\right)$ plane was chosen to be $2.5\times10^{-3}$ a.u., $1.25\times10^{-3}$ a.u., and $6.25\times10^{-4}$ a.u. at the wavelengths of 800 nm, 1200 nm, and 1600 nm, respectively. With the number of trajectories mentioned above $\left(\sim10^{9}\right)$ this ensures the convergence of the photoelectron energy spectrum up to $10^{-5}$ of its maximum value. 

We benchmark our present SCTS model against the exact numerical solution of the time-dependent Schr\"odinger equation (TDSE) and also compare with results of the previous QTMC model.

In order to numerically solve the TDSE
\begin{equation}
\label{TDSE}
i\frac{\partial |\psi(t)\rangle}{\partial t}=\left\{-\frac{\Delta}{2}+V(r)+zF(t)\right\}|\psi(t)\rangle
\end{equation}
in the dipole approximation for a single active electron, we employ the generalized pseudo-spectral method \cite{Tong97,Tong00,Tong05}. This method combines the discretization of the radial coordinate optimized for the Coulomb singularity with quadrature methods to allow stable long-time evolution using a split-operator representation of the time-evolution operator. Both the bound as well as the unbound parts of the wave function $|\psi(t)\rangle$ can be accurately represented. The atomic potential $V(r)$ is taken to be the Coulomb potential, $V(r)=-1/r$. Propagation of the wave function is started from the ground state of hydrogen. Due to the cylindrical symmetry of the system the magnetic quantum number $m=0$ is conserved. After the end of the laser pulse the wave function is projected on eigenstates $|p,\ell\rangle$ of the free atomic Hamiltonian with positive eigenenergy $E=p^2/2$ and orbital quantum number $\ell$ to determine the transition probabilities $R(\vec p)$ to reach the final state $|\phi_{\vec p}\rangle$ (see Refs.~\cite{Schoeller86,Mess65,Dionis97}):
\begin{equation}
R(\vec p)=\frac{1}{4\pi p}\left\vert \sum_{l}e^{i\delta
_{\ell}(p)}\
\sqrt{2l+1}P_{\ell}(\cos \theta )\left\langle p,\ell\right. \left\vert \psi
(t_{f})\right\rangle \right\vert ^{2}.  \label{coulomb}
\end{equation}
In Eq.~(\ref{coulomb}), $\delta _{\ell}(p)$ is the momentum-dependent atomic phase shift, $\theta $ is the angle between the electron momentum $\vec{p}$ and the polarization direction of the laser field $\vec{e}_{z}$ and $P_{\ell}$ is the Legendre polynomial of degree $\ell$. In order to avoid unphysical reflections of the wave function at the boundary of the system, the length of the computing box was chosen to be 1200 a.u.\ ($\sim 65$ nm) which is much larger than the maximum quiver amplitude $\alpha=F_0/\omega^2=62$ a.u. at the intensity of $0.9\times10^{14}$ W/cm$^2$ and the wavelength of 1600 nm. The maximum angular momentum included was $\ell_{\max}=300$.

\begin{figure}[h]
\begin{center}
\includegraphics[width=.8\textwidth]{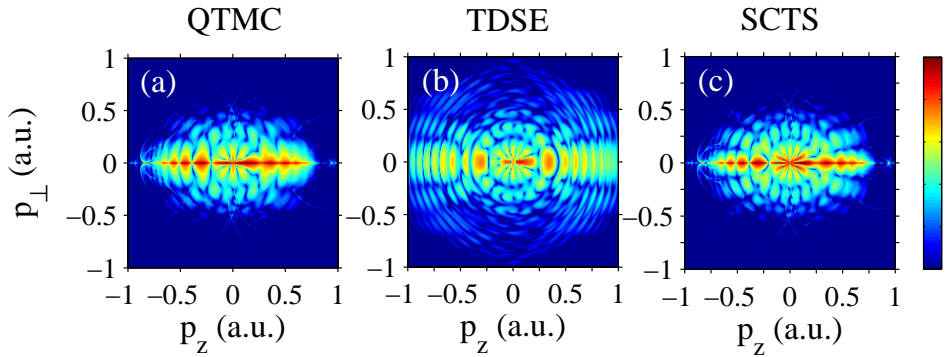} 
\end{center}
\caption{(Color online) Vectorial momentum distributions for the H atom ionized by a laser pulse with a duration of $n=8$ cycles, wavelength of $\lambda=800$ nm, and peak intensity of $I=0.9\times10^{14}$ W/cm$^2$ obtained from (a) the QTMC model, (b) solution of the TDSE, and (c) the present SCTS model. The distributions are normalized to the total ionization yield. A logarithmic color scale is used. The laser field is linearly polarized along the $z$-axis. A logarithmic color scale is used spanning 5 orders of magnitude from the highest intensity (dark red) to lowest intensity (blue).}   
\label{fig1}
\end{figure}

\begin{figure}[h]
\begin{center}
\includegraphics[width=.8\textwidth]{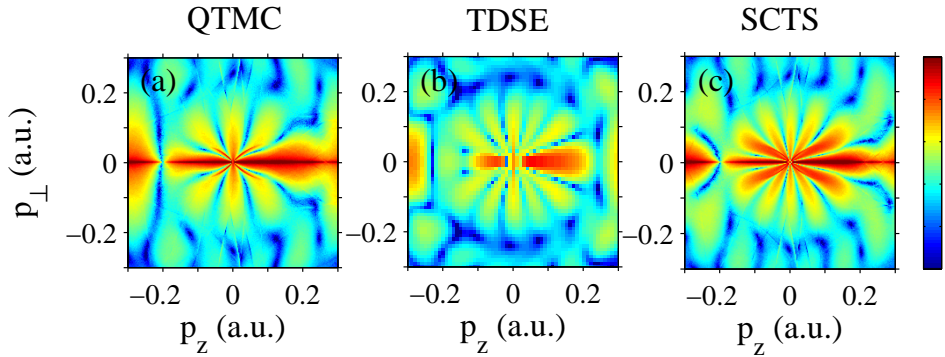} 
\end{center}
\caption{(Color online) Magnification of Fig.\ \ref{fig1} for $|p_z|,|p_\perp|<0.3$ a.u.}
\label{fig2}
\end{figure}

\begin{figure}[h]
\begin{center}
\includegraphics[width=0.8\textwidth]{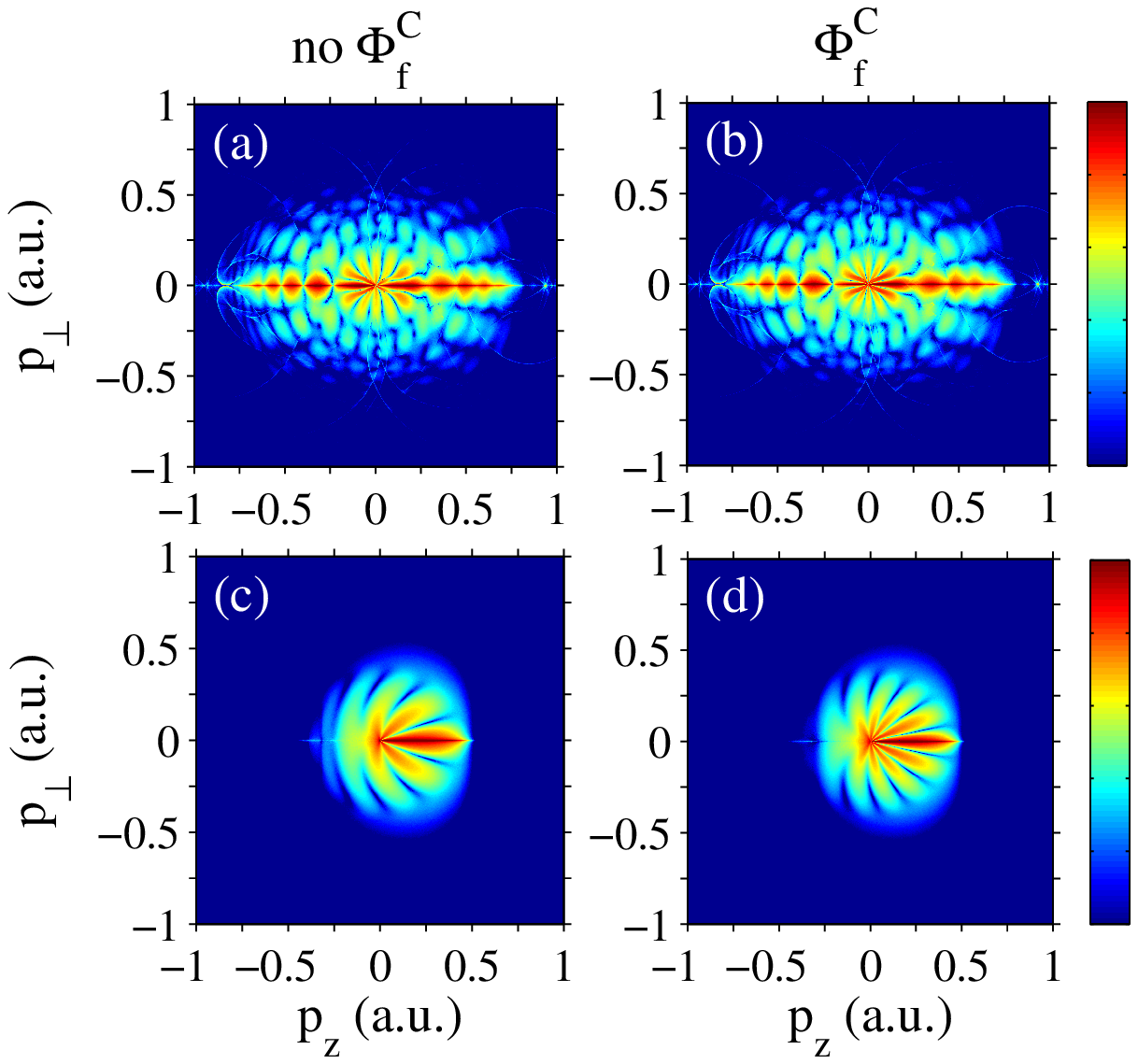} 
\end{center}
\caption{(Color online) Vectorial momentum distribution from the present SCTS model for low-energy electrons without (panels a, c) and with (b, d) inclusion of the post-pulse Coulomb phase $\Phi_f^C(\tau_f)$ [Eq.~(\ref{phase_asym})] for the eight-cycle pulse of Fig.~\ \ref{fig1} (a, b) and a single-cycle pulse (c, d) with all other laser parameters identical. The distributions are normalized to the total ionization yield. A logarithmic color scale is used spanning 5 orders of magnitude from the highest intensity (dark red) to lowest intensity (blue).}
\label{fig3}
\end{figure}

We first turn our attention to the vectorial photoelectron momentum distribution in the $(p_z,p_\perp)$ plane (Fig.~\ref{fig1}).
For these semiclassical simulations we employ the initial distribution [Eq.~(\ref{tunrate})] with zero initial parallel velocity. Overall, as noted previously, the momentum distributions of the SCTS and QTMC models, both of which calculated from the same initial distributions, qualitatively resemble the TDSE results quite well. However, a close-up of the low-energy spectrum (Fig.~\ref{fig2}) shows marked deviations.
For $|p|\leq 0.3$ a.u.~and energies well below $U_p=0.2$ a.u., the vectorial momentum distribution displays a fan-like interference structure similar to that of Ramsauer-Townsend difraction oscillations \cite{Arbo06,Moshammer09,Arbo062, Arbo08}. The number of radial nodal lines is controlled by the dominant partial-wave angular momentum $\ell_c$ in Eq.~(\ref{coulomb}), i.e., $R(\vec{p}) \sim |P_{l_c}(cos \theta)|^2$ (see Refs.~\cite{Arbo062, Arbo08}). While the SCTS model closely matches the nodal pattern of the TDSE, the QTMC model yields fewer nodal lines, which is a direct consequence of the underestimate of the Coulomb interaction in the QTMC treatment of the interference phase. This effect of neglecting the elastic scattering in the Coulomb field occurs both during the laser pulse [Eq.~(\ref{phase_decom})] and after [Eq.~(\ref{phase_asym})]. The magnitude of the latter is illustrated in Fig.~\ref{fig3} where we display the effect of $\Phi_{f}^{C}(\tau_f)$ for both an ultrashort single-cycle pulse and the longer eight-cycle pulse. The post-pulse Coulomb phase is more pronounced for shorter $\tau_f$ as the electron is still closer to the nucleus at the end of the pulse.

\begin{figure}[h]
\begin{center}
\includegraphics[width=0.8\textwidth]{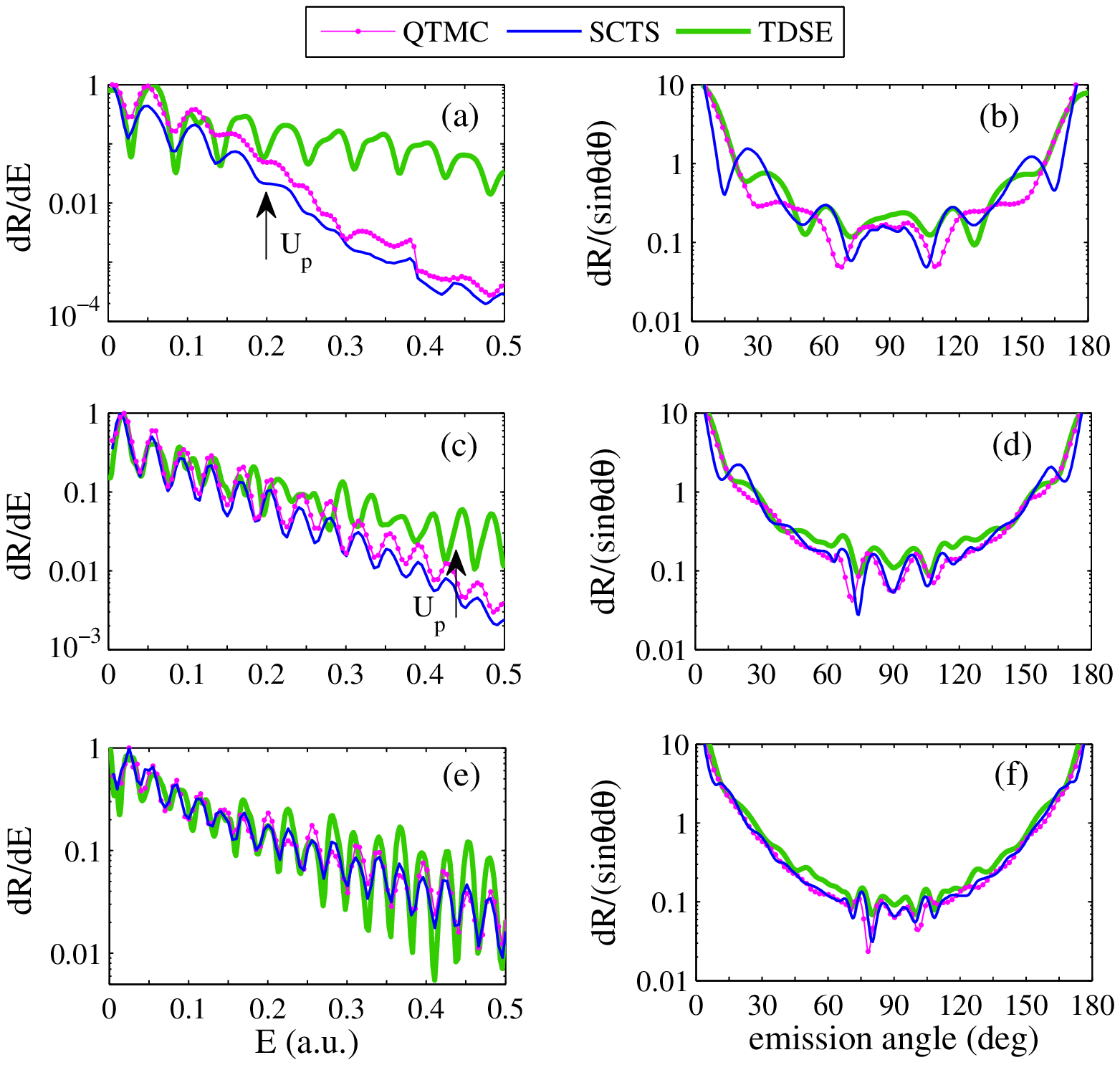} 
\end{center}
\caption{(Color online) Energy spectra [panels (a), (c), and (e)] and angular distributions [panels (b), (d), and (f)] of the photoelectrons for ionization of H at an intensity of $9\times10^{13}$ W/cm$^2$ and a pulse duration of 8 cycles obtained from the QTMC model [thin (magenta) curve with solid circles], the SCTS model [solid (blue) curve], and TDSE [thick (green) curve]. The distributions [(a),(b)], [(c),(d)], and [(e),(f)] correspond to the wavelengths of 800, 1200, and 1600 nm, with Keldysh parameters of 1.12, 0.75, and 0.56, respectively. The energy spectra are normalized to the peak value, the angular distributions are normalized to the total ionization yield and show the spectrum for electrons with asymptotic energies $E<U_p$. The energy equal to $U_{p}$ is shown by arrows in panels (a) and (c).}
\label{fig4}
\end{figure}

\begin{figure}[h]
\begin{center}
\includegraphics[width=0.45\textwidth]{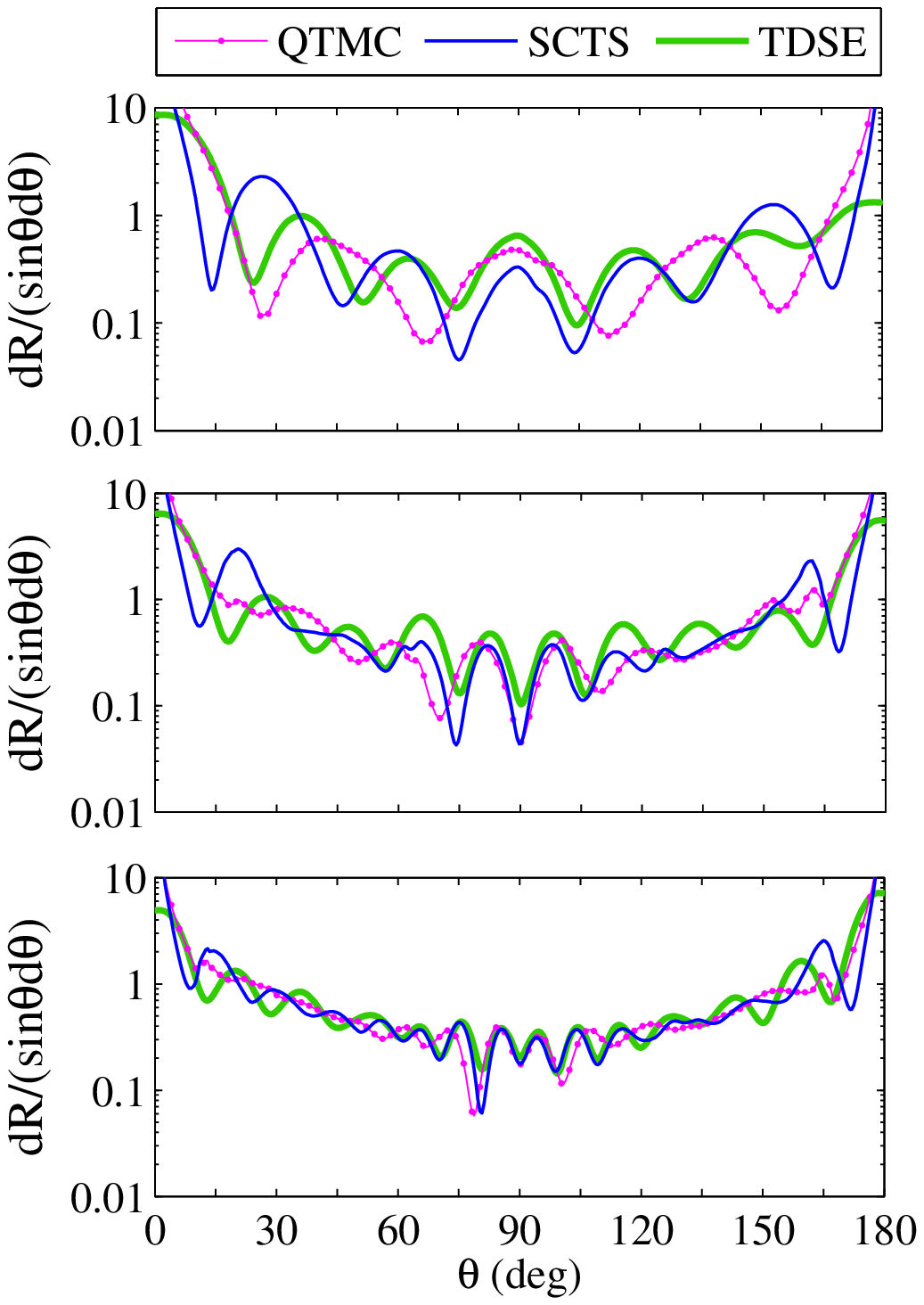} 
\end{center}
\caption{(Color online) Angular distributions for low-energy electrons (innermost fan-like structure, c.f.\ Fig.\ \ref{fig2}): (a) $E<0.022$ a.u. for $\lambda$=800 nm, (b) $E<0.031$ a.u. for $\lambda$=1200 nm, and (c) $E<0.036$ a.u. for $\lambda=1600$ nm. Cut-off energies have been determined from TDSE results.}
\label{fig5}
\end{figure}

For a quantitative comparison of different methods we consider the singly-differential angular distribution (Fig.~\ref{fig4})
\begin{equation}
\frac{dR}{\sin\theta d\theta}=2\pi\int_0^\infty dE\sqrt{2E} \,R[\vec p(E)]
\end{equation}
and the photoelectron spectrum
\begin{equation}
\frac{dR}{dE}=2\pi\sqrt{2E}\int_0^\pi d\theta \sin\theta \, R[\vec p(\theta)]\, .
\end{equation}
The energy spectra feature pronounced ATI peaks. These are qualitatively reproduced by the semiclassical methods. However, only for the low-order peaks the semiclassical approximation can quantitatively reproduce the amplitude of the oscillations \cite{Arbo10, Arbo14}. This is closely related to the fact that the initial conditions from the tunneling step [Eq.~(\ref{tunrate})] provide too few trajectories with large longitudinal momenta that could account for intercycle interferences, the semiclassical origin of the ATI modulation at large momenta. For the same reason the photoelectron spectrum $dR/dE$ falls off too rapidly for energies exceeding $\sim U_p$. The semiclassical angular distributions reproduce the Ramsauer-Townsend diffraction oscillations \cite{Arbo062, Arbo08}. The modulation amplitude as well as the position of the minima of the SCTS agree better with the TDSE compared to the QTMC model because of the improved interference phase. The difference is more pronounced for the angular distribution of low-energy electrons (Fig.~\ref{fig5}).

Obviously, further improvement of the semiclassical description of the energy and angular distributions of photoelectrons require an amended initial distribution emerging from the tunneling step. In order to improve the initial conditions for the propagation of classical trajectories, we set the initial parallel velocity $v_{0,z}$ at every ionization time $t_0$ in Eq.~(\ref{tunrate}) to a nonzero value predicted by the strong-field approximation (see Refs.~\cite{Yan10, Popruzhenko08, Popruzhenko08PRA, Li16}). For the pulse defined in Eq.~(\ref{vecpot}) it can be approximated as
\begin{equation}
v_{z,0}\left(t_0\right)=-\frac{1}{c}A_{z}\left(t_{0}\right)\left[\sqrt{1+\gamma^{2}\left(t_0,v_{0,\perp}\right)}-1\right],
\label{vz0sfa}
\end{equation} 
where 
\begin{equation}
\gamma\left(t_0,v_{0,\perp}\right)=\frac{\omega\sqrt{2I_{p}+v_{0,\perp}^{2}}}{F_0\sin^2\left(\frac{\omega t_0}{2n_p}\right)\left|\cos\left(\omega t_0\right)\right|}
\label{gameff}
\end{equation}
is the effective Keldysh parameter \cite{Li16}. In the tunneling limit $\gamma\left(t_0,v_{0,\perp}\right)\to 0$ the longitudinal initial velocity $v_{z,0}(t_0)$ vanishes.

\begin{figure}[h]
\begin{center}
\includegraphics[width=0.8\textwidth]{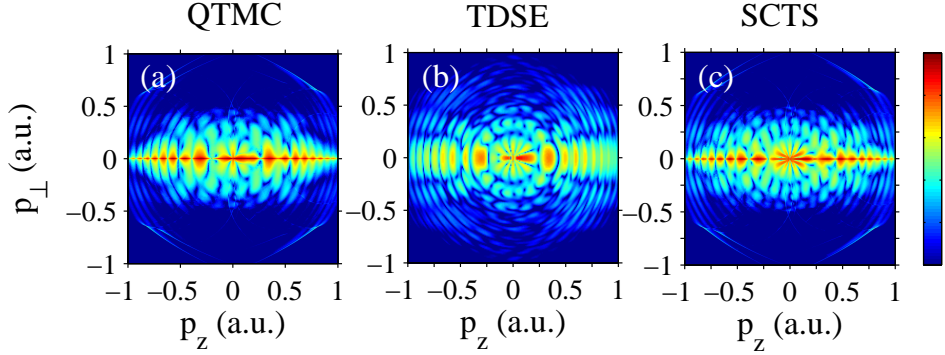} 
\end{center}
\caption{(Color online) Same as Fig.\ \ref{fig1} but with nonzero initial parallel velocity.}
\label{fig6}
\end{figure}

\begin{figure}[h]
\begin{center}
\includegraphics[width=0.8\textwidth]{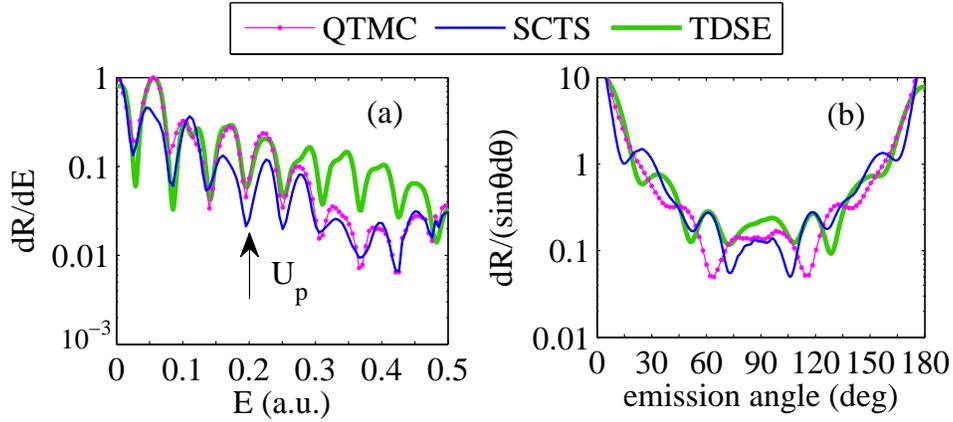} 
\end{center}
\caption{(Color online) Same as Fig.\ \ref{fig4} for $\lambda =  800$ nm with nonzero initial parallel velocity.}
\label{fig7}
\end{figure}

Employing Eq.~(\ref{vz0sfa}) as initial condition for CTMC trajectories taking off at $t_0$ at the tunneling exit yields improved agreement between the SCTS model and the TDSE for both the vectorial momentum distribution (Fig.~\ref{fig6})
and the singly differential distributions $dR/dE$ and $dR/\left(\sin\theta d\theta\right)$ (Fig.~\ref{fig7}).
Indeed, the SCTS model can now better reproduce the energy spectrum obtained from the TDSE, see Fig.~\ref{fig7}(a). For angular distribution the agreement between the QTMC and the TDSE worsens whereas the agreement between the SCTS and the TDSE improves [compare Fig.~\ref{fig7}(b) with Fig.~\ref{fig4}(b)]. These results clearly suggest that the main source of deviations of the SCTS model from the TDSE are the errors in treating the tunneling step rather than the semiclassical description of the post-tunneling propagation.
 
\section{Conclusions and outlook}\label{sec5}
We have developed a semiclassical two-step model for strong-field ionization that describes quantum interference and accounts for the Coulomb potential beyond the semiclassical perturbation theory. In the SCTS model the phase associated with every classical trajectory is calculated using the semiclassical expression for the matrix element of the quantum mechanical propagator. For identical initial conditions after the tunneling ionization step taken from standard tunnel ionization rates \cite{KrainovBook}, the SCTS model yields closer agreement with the exact solution of the Schr\"odinger equation than the previously proposed QTMC model. Furthermore, after improving the input from the tunneling ionization step by including nonzero parallel velocities in the initial conditions for the motion after tunneling, the SCTS model yields significantly improved agreement in the angular distribution, i.e., the position of interference fringes with the TDSE results. Remaining differences in the intensity of energy distributions are traced back to improvable starting conditions (in particular the choice of parallel velocities) of classical trajectories.

The present SCTS model can be extended to multielectron targets in a straightforward fashion by the inclusion of dynamical Stark shifts and polarization-induced dipole potentials. Semiclassical models of this type will allow to investigate the role of the multielectron polarization effect in the formation of the interference structure in the electron momentum distributions. Since the multielectron potential affects both the exit point and the electron dynamics in the continuum, pronounced imprints of the polarization effects in the interference patterns are expected. Finally, the two-step semiclassical models accounting for both the interference and multielectron effects can provide a valuable tool for investigation of the delays in photoemission, which is presently one of the most intensively studied problems in strong-field physics and attosecond science.  

\acknowledgments
We are grateful to Mr.~Janne Solanp\"a\"a (Tampere University of Technology) for stimulating discussions. This work has been supported by the European Community's FP7 through the CRONOS Project No.~280879, The Academy of Finland Project No.~267686, the Nordic Innovation through its Top-Level Research Initiative Project No.~P-13053, the COST Action CM-1204 (XLIC-XUV/X-ray light and fast ions for ultrafast chemistry), the STSM Grant from the COST Action CM-1204, an ERC-StG (Project No.~277767, TDMET), the VKR Center of Excellence QUSCOPE, FWF-SFB049 NextLite, CONICET (Argentina PIP0386), ANPCyT (Argentina PICT-2014-2363), and UBACyT0617BA.

\end{document}